\documentclass[11pt,a4paper]{article}
\usepackage[hyperref]{naaclhlt2019}
\usepackage{times}
\usepackage{latexsym}
\usepackage{bbm}
\usepackage{url}
\usepackage{epsfig}
\usepackage{graphicx}
\usepackage{amsmath}
\usepackage{amssymb,amsthm}
\usepackage{multirow}
\usepackage{enumitem}
\usepackage{url}
\usepackage{verbatim}
\usepackage{changepage}
\usepackage{booktabs}
\usepackage{xcolor}
\newcommand{\eat}[1]{}
\newcommand{\tightlist}{\itemsep=-2pt}

\newcommand{\ie}{{\em i.e.}}

\aclfinalcopy 
\def\aclpaperid{***} 

\newtheorem*{problem}{Problem Statement}

\title{{OpenKI: Integrating Open Information Extraction and Knowledge Bases \\with Relation Inference}}

\author{Dongxu Zhang$^1$\thanks{~~~This work was performed while at Amazon.}, Subhabrata Mukherjee$^2$, Colin Lockard$^2$,\\
      {\bf Xin Luna Dong$^2$, \bf Andrew McCallum$^1$}
      \\
      $^1$University of Massachusetts Amherst\\
      \url{{dongxuzhang,mccallum}@cs.umass.com}\\ 
      $^2$Amazon\\
      \url{{subhomj,clockard,lunadong}@amazon.com}\\
      }

\date{}

\begin{document}
\maketitle

\begin{abstract}

In this paper, we consider advancing web-scale knowledge extraction and alignment by integrating OpenIE extractions in the form of (subject, predicate, object) triples with Knowledge Bases (KB). Traditional techniques from universal schema and from schema mapping fall in two extremes: either they perform instance-level inference relying on embedding for (subject, object) pairs, thus cannot handle pairs absent in any existing triples; or they perform predicate-level mapping and completely ignore background evidence from individual entities, thus cannot achieve satisfying quality.

 We propose \emph{OpenKI} to handle sparsity of OpenIE extractions {by performing instance-level inference}: for each entity, we encode the rich information in its neighborhood in both KB and OpenIE extractions, and leverage this information in relation inference by exploring different methods of aggregation and attention. {In order to handle unseen entities, our model is designed without creating entity-specific parameters.} Extensive experiments show that this method not only significantly improves state-of-the-art for conventional OpenIE extractions like ReVerb, but also boosts the performance on OpenIE from semi-structured data, where new entity pairs are abundant and data are fairly sparse.

\end{abstract}

\section{Introduction}


Web-scale knowledge extraction and alignment has been a vision held by different communities for decades. The Natural Language Processing (NLP) community has been focusing on knowledge extraction from texts. They apply either closed information extraction according to an ontology~\cite{mintz2009distant,guodong2005exploring}, restricting to a subset of relations pre-defined in the ontology, or open information extraction (OpenIE) to extract free-text relations~\cite{banko2007open,DBLP:conf/emnlp/FaderSE11}, leaving the relations unaligned and thus potentially duplicated. The Database (DB) community has been focusing on aligning relational data or WebTables~\cite{cafarella2008webtables} by schema mapping~\cite{rahm2001survey}, but the quality is far below adequate for assuring correct data integration.

We propose advancing progress in this direction by applying {\em knowledge integration} from OpenIE extractions. OpenIE extracts SPO (subject, predicate, object) triples, where each element is a text phrase, such as {\bf E1:} {\em (``Robin Hood'', ``Full Cast and Crew'', ``Leonardo Decaprio")} and {\bf E2}: {\em (``Ang Lee", ``was named best director for", ``Brokeback")}. OpenIE has been studied for text extraction extensively~\cite{DBLP:conf/naacl/YatesBBCES07,DBLP:conf/emnlp/FaderSE11,DBLP:conf/emnlp/MausamSSBE12}, 
and also for semi-structured sources~\cite{bronzi2013extraction}, thus serves an effective tool for web-scale knowledge extraction. The remaining problem is to align text-phrase predicates\footnote{We also need to align text-phrase entities, which falls in the area of entity linking~\cite{dredze2010entity,ji2014overview}; it is out of scope of this paper and we refer readers to relevant references.} from OpenIE to knowledge bases (KB). Knowledge integration answers the following question: given an OpenIE extraction $(s, p, o)$, how can one populate an existing KB using relations in the pre-defined ontology? 

The problem of knowledge integration is not completely new. The DB community has been solving the problem using schema mapping techniques, identifying mappings from a source schema (OpenIE extractions in our context) to a target schema (KB ontology in our context)~\cite{rahm2001survey}. Existing solutions consider {\em predicate-level} (\ie, attribute) similarity on names, types, descriptions, instances, and so on, and generate mappings like ``email" mapped to ``email-address"; ``first name" and ``last name" together mapped to ``full name". However, for our example ``Full Cast and Crew", which is a union of multiple KB relations such as ``directed\_by", ``written\_by", and ``actor", it is very hard to determine a mapping at the predicate level.

On the other hand, the NLP community has proposed {\em Universal Schema}~\cite{riedel2013relation} to apply {\em instance-level} inference from both OpenIE extractions and knowledge in existing knowledge bases: given a set of extractions regarding an entity pair $(s, o)$ and also information of each entity, infer new relations for this pair. 
One drawback of this method is that it cannot handle unseen entities and entity pairs. Also, the technique tends to overfit when the data is sparse due to large number of parameters for entities and entity pairs. Unfortunately, in the majority of the real extractions we examined in our experiments,  
we can find only 1.4 textual triples on average between the subject and object. The latest proposal {\em Rowless Universal Schema}~\cite{verga-neelakantan-mccallum:2017:EACLlong} removes the entity-specific parameters and makes the inference directly between predicates and relations, thereby allowing us to reason about unseen entity pairs. However, it completely ignores the entities themselves, so in a sense falls back to {\em predicate-level} decisions, especially when only one text predicate is observed.

In this paper we propose a solution that leverages information about the individual entities whenever possible, and falls back to predicate-level decisions only when both involved entities are new. Continuing with our example {\bf E1} -- if we know from existing knowledge that ``Leonardo" is a famous actor and has rarely directed or written a movie, we can decide with a high confidence that this predicate maps to @film.actor in this triple, even if our knowledge graph knows nothing about the new movie ``Robin Hood". In particular, we make three contributions in this paper. 
\begin{enumerate}[leftmargin=*]\tightlist
    \item We design an embedding for each entity by exploring rich signals from its neighboring relations and predicates in KB and OpenIE. This embedding provides a soft constraint on which relations the entities are likely to be involved in, while keeping our model free from creating new entity-specific parameters so allowing us to handle unseen entities during inference.
    \item Inspired by predicate-level mapping from schema mapping and instance-level inference from universal schema, we design a joint model that leverages the neighborhood embedding of entities and relations with different methods of aggregation and attention. 
    \item Through extensive experiments on various OpenIE extractions and KB, we show that our method improves over state-of-the-arts by 33.5\% on average across different datasets.
\end{enumerate}

In the rest of the paper, we define the problem formally in Section~\ref{sec:def_and_overview}, present our method in Section~\ref{sec:model}, describe experimental results in Section~\ref{sec:experiment}, and discuss related work in Section~\ref{sec:related_work}.

\section{Problem Overview}
\label{sec:def_and_overview}


\begin{problem}
Given (i) an existing knowledge base $KB$ of triples $(s,p,o)$ -- where  $s, o \in E^{KB}$ (set of KB entities) and $ p \in R^{KB}$ (set of KB relations), and (ii) a set of instances $(s',p',o')$ from OpenIE extraction ($s'$ and $o'$ may not belong to $E^{KB}$, and $p'$ are text predicates)\footnote{ In this paper, a `relation' always refers to a KB relation, whereas a `predicate' refers to an OpenIE textual relation.}: predict $score(s',p, o')$ -- where $p \in R^{KB}$.

\end{problem}

\noindent For example, given {\bf E1} and {\bf E2} as OpenIE extractions and background knowledge bases (KB) like IMDB, we want to predict ``@film.actor" relation  given E1 and ``@film.directed\_by" relation given E2 as the target KB relations between the participating entities. 
Particularly, we want to perform this relation inference at {\em instance-level}, which can be different for different entities sharing the same predicate. Table~\ref{tab:notation} introduces important notations used in this paper. 


\begin{table}[h]
\small
\begin{adjustwidth}{-1.0em}{}
\begin{tabular}{l|l} 
\toprule
$s$ & Subject\\
$o$ & Object\\
$p$ & KB relation or text predicate \\
$v_s, v_p, v_o, v_{s,o}$ & Embedding vectors of $s,p,o$ and $(s,o)$ \\
$S(s,p,o)$ & Scoring function for $(s,p,o)$ to be true\\
$Agg_{p \in R(s,o)} (v_p)$ & Aggregation function over embeddings\\
& ($v_p$) of $p$ shared by $s$ and $o$. \\
\bottomrule
\end{tabular}
\caption{\label{tab:notation} Notation table.}
\end{adjustwidth}{}{}
\vspace{-1em}
\end{table}

\vspace{-1em}

\subsection{Existing Solution and Background}
\label{sec:background}

\noindent {\bf Universal Schema (F-Model)}~\cite{riedel2013relation} is modeled as a matrix \textbf{\underline{f}}actorization task where entity pairs, e.g., {\em (RobinHood, Leonardo Decaprio)} form the rows, and relations from OpenIE and KB form the columns (e.g., @film.actor, ``Full Cast and Crew"). During training, we observe some positive entries in the matrix and the objective is to predict the missing cells at test time. Each (subject, predicate, object) triple is scored as:
\begin{equation}
\label{eq:universal_schema}
S_{F}(s, p, o) = v_{s,o} \cdot v_{p}^{T} \nonumber 
\end{equation}
where, $v_{s,o} \in \mathcal{R}^d$ is the embedding vector of the entity pair (subject, object), 
$v_{p}$ is the embedding vector of a KB relation or OpenIE predicate, and the triple score is obtained by their dot product. The parameters $v_{p}$ and $v_{s,o}$ are randomly initialized and learned via gradient descent.

One of the drawbacks of universal schema is the explicit modeling of entity pairs using free parameters $v_{s,o}$. Therefore, it cannot model unseen entities. This also makes the model overfit on our data as the number of OpenIE text predicates observed with each entity pair is rather small (1.4 on average in our datasets). 

\noindent{\bf Universal Schema (E-Model) }~\cite{riedel2013relation} considers \textbf{\underline{e}}ntity-level information, thus decomposing the scoring function from the F-model as follows:
\begin{align}
\label{eq:E}
S_{E}(s, p, o) &= S_{subj}(s,p) + S_{obj}(p,o) \nonumber \\
&= v_{s} \cdot v^{subj^T}_{p} + v_{o} \cdot v^{obj^T}_{p}
\end{align}
where each relation is represented by two vectors corresponding to its argument type for a subject or an object. The final score is an additive summation over the subject and object scores $S_{subj}$ and $S_{obj}$ that implicitly contain the argument type information of the predicate $p$. Thus, a joint F- and E-model of $S_{F+E} = S_F + S_E$ can perform relation inference at {\em instance-level} considering the entity information. Although the {E-model} captures rich information about entities, it still cannot deal with unseen entities due to the entity-specific free parameters $v_s$ and $v_o$.

\noindent {\bf Rowless Universal Schema (Rowless)}~\cite{verga-neelakantan-mccallum:2017:EACLlong} handles new entities as follows. 
It considers all relations in KB and OpenIE that the subject $s$ and object $o$ co-participates in (denoted by $R(s,o)$), and represents the entity pair with an aggregation over embeddings of these relations.
\begin{align}
\label{eq:rowless_uschema}
& v^{Rowless}_{s,o}  = Agg_{p' \in R(s,o)}(v_{p'}) \nonumber \\
& S_{Rowless}(s, p, o) = v^{Rowless}_{s,o} \cdot v_{p}^{T}
\end{align}
$Agg(.)$ is an aggregation function like average pooling, max pooling, hard attention (Rowless MaxR) or soft attention given query relations (Rowless Attention)~\cite{verga-neelakantan-mccallum:2017:EACLlong}.
The Rowless model ignores the individual information of entities, and therefore falls back to making {\em predicate-level} decisions in a sense, especially when there are only a few OpenIE predicates for an entity pair.
 

\begin{figure*}
	\centering
	\includegraphics[width=0.95\textwidth]{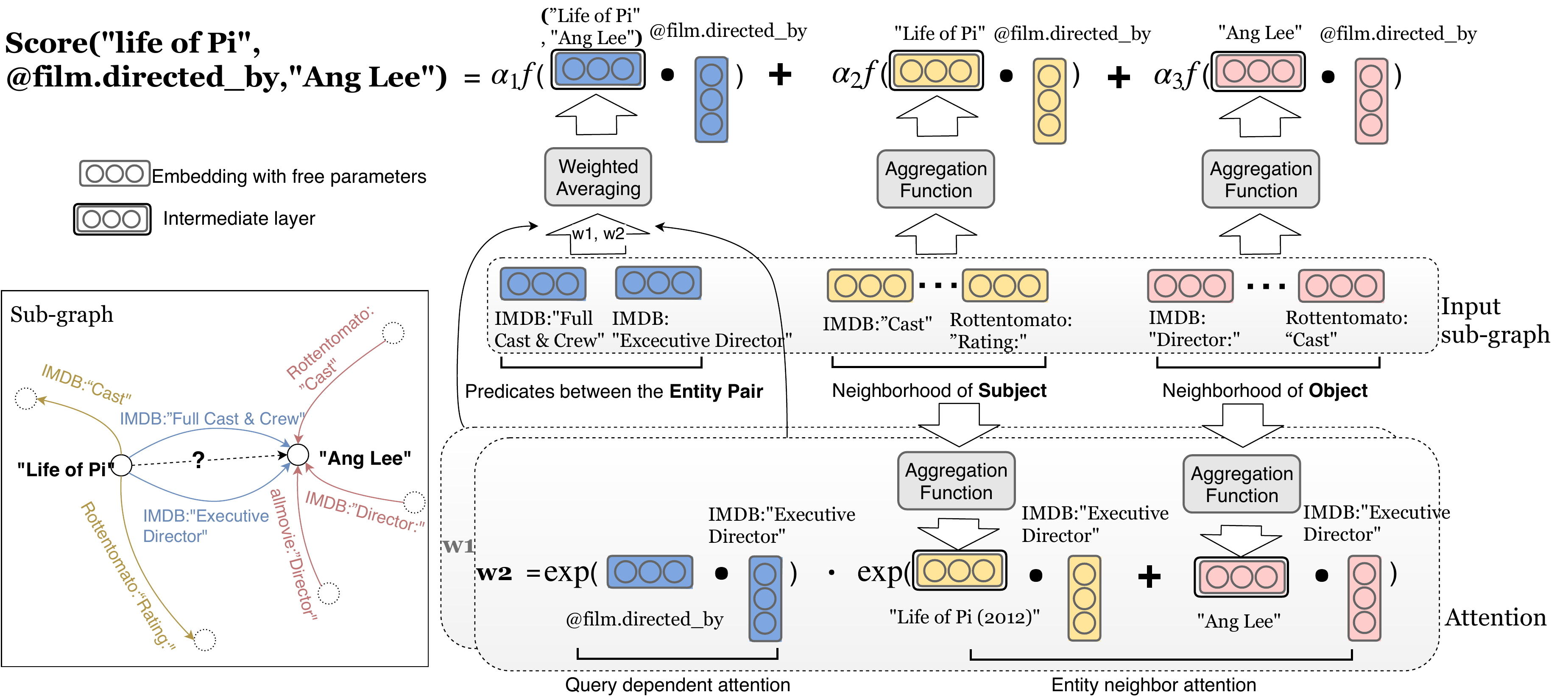}
	\caption{\small Architecture of the proposed method. In this example, the ENE model uses ``Ang Lee's'' neighboring predicates ``{IMDB:Director"} and ``{allmovie:Director}" for predicting the target KB relation ``@film.directed\_by". The attention mechanism assigns a larger weight over ``IMDB:Executive Director'' for generating entity pair embedding. Different colors of vectors represent different sets of parameters. The Entity Neighborhood Encoder (ENE) (yellow and pink) model contributes the following components to the final scoring function: (1) entity neighborhood scores $S^{ENE}_{subj}$ and $S^{ENE}_{obj}$ of the subject and object respectively; (2) query and neighborhood signal for the attention module to calculate the weight of each text predicate (Blue) and the attention score $S_{Att}(subj,pred,obj)$. }
	\label{fig:modelstructure}
	\centering
\end{figure*}

\section{Our Approach}
\label{sec:model}

We propose \emph{OpenKI} for {\em instance-level} relation inference such that it (i) captures rich information about each entity from its neighborhood KB relations and text predicates to serve as background knowledge and generalizes to unseen entities by {\em not} learning any entity-specific parameters (only KB relations and OpenIE predicates are parameterized)
{(ii) considers both shared predicates and entity neighborhood information to encode entity pair information. 
Figure~\ref{fig:modelstructure} shows the architecture of our model.

\subsection{Entity Neighborhood Encoder (ENE)}

The core of our model is the Entity Neighborhood Encoder. Recall that
Rowless Universal Schema represents each entity pair with common relations shared by this pair. However, it misses critical information when entities do not only occur in the current entity pair, but also interact with other entities. This \emph{entity neighborhood} can be regarded as a soft and fine-grained entity type information that could help infer relations when observed text predicates are ambiguous (polysemous), noisy (low quality of data source) or low-frequency (sparsity of language representation).
\footnote{{Note that, the notion of \emph{entity neighborhood} is different from the \emph{Neighborhood} model in the Universal Schema work~\cite{riedel2013relation}. Our \emph{entity neighborhood} captures information of each entity, whereas their {Neighborhood} model leverages prediction from similar predicates.}}

Our aim is to incorporate this entity neighborhood information into our model for {\em instance-level} relation inference while keeping it free of entity-specific parameters. 
To do this, for each entity, we leverage all its {\em neighboring KB relations and OpenIE predicates} for relation inference. 
We aggregate their embeddings to obtain two scores for the subject and object separately in our ENE model. The subject score $S^{ENE}_{subj}$ for an entity considers the aggregated embedding of its participating KB relations and OpenIE predicates where it serves as a subject (similar for the object score $S^{ENE}_{obj}$):

\begin{align}
\nonumber
&v^{agg}_{subj} = Agg_{p' \in R(s,\cdot)}(v^{subj}_{p'})\\
\nonumber
&S^{ENE}_{subj}(s, p) = v^{agg}_{subj} \cdot v^{subj\ T}_{p}\\
\nonumber
&v^{agg}_{obj} = Agg_{p' \in R(\cdot,o)}(v^{obj}_{p'})\\
\label{eq:E_rowless}
&S^{ENE}_{obj}(p, o) = v^{agg}_{obj}  \cdot v^{obj\ T}_{p}
\end{align}
$R(s, .)$ denotes all neighboring relations and predicates of the subject $s$ (similar for the object).  $v_p^{subj}$ and $v_p^{obj}$ are the only free parameters in ENE. These are randomly initialized and then learned via gradient descent. We choose average pooling as our aggregation function to capture the proportion of different relation and predicate types within the target entity's neighborhood. 

\subsection{Attention Mechanism}

Given multiple predicates between a subject and an object, only some of them are important for predicting the target KB relation between them. For example, in Figure~\ref{fig:modelstructure}, the predicate ``Executive Director" is more important than ``Full Cast \& Crew" to predict the KB relation ``@film.directed\_by" between ``Life of Pi" and ``Ang Lee".  

We first present a query-based attention mechanism from earlier work, and then present our own solution with a neighborhood attention and combining both in a dual attention mechanism.

\subsubsection {Query Attention} 
The first attention mechanism uses a query relation $q$ (i.e., the target relation we may want to predict) to find out the importance (weight) $w_{p|q}$ of different predicates $p$ with respect to $q$ with $v_p$ and $v_q$ as the corresponding relation embeddings.
\begin{align}
& w_{p|q} = \frac{\exp(v_q \cdot v^T_{p})}{\sum_{p'} \exp(v_q \cdot v^T_{p'})} \nonumber 
\end{align}
Thus, given each query relation $q$, the model tries to find evidence from predicates that are most relevant to the query. Similar techniques have been used in~\cite{verga-neelakantan-mccallum:2017:EACLlong}.
We can also use hard attention (referred as MaxR) instead of soft attention where the maximum weight is replaced with one and others with zero.
One potential shortcoming of this attention mechanism is its sensitivity to noise, whereby it may magnify sparsely observed predicates between entities.

\subsubsection {Neighborhood Attention} 

In this attention mechanism, we use the subject and object's neighborhood information as a filter to remove unrelated predicates. Intuitively, the entity representation generated by the ENE from its neighboring relations can be regarded as a soft and  fine-grained entity type information. 

Consider the embedding vectors $v^{agg}_{subj}$ and $v^{agg}_{obj}$ in Equation~\ref{eq:E_rowless} that are aggregated from the entity's neighboring predicates and relations using an aggregation function. We compute the similarity $w_{p|Nb}$ between an entity's neighborhood information given by the above embeddings and a text predicate $p$ to enforce a soft and fine-grained argument type constraint over the text predicate:
\begin{align}
& w_{p|Nb} = \frac{\exp(v^{agg}_{subj} \cdot v^{subj^T}_{p} + v^{agg}_{obj} \cdot v^{obj^T}_{p})}{\sum_{p'} \exp(v^{agg}_{subj} \cdot v^{subj^T}_{p'} + v^{agg}_{obj} \cdot v^{obj^T}_{p'})} \nonumber \nonumber
\end{align}

Finally, we combine both the  query-dependent and neighborhood-based attention into a \emph{Dual Attention} mechanism:
\begin{align}
& w_{p|q+Nb} = w_{p|q} \cdot w_{p|Nb} \nonumber \\
& w_p  =  \frac{w_{p|q+Nb}}{\sum_{p'} w_{p'|q+Nb}}\nonumber
\end{align}
And the score function is given by:
\begin{align}
\nonumber
S^{Att}(s,q,o) &= Agg_{p \in R(s,o)}(v_{p}) \cdot v_q^T\\
\label{eq:attscore}
 &=\big(\sum_{p} w_{p} v_{p} \big)  \cdot v_q^T
\end{align}

\subsection{Joint Model: OpenKI}


All of the above models capture different types of features. Given a target triple $(s,p,o)$, we combine scores from Eq.~\ref{eq:E_rowless} and Eq.~\ref{eq:attscore} in our final OpenKI model. It aggregates the neighborhood information of $s$ and $o$ and also uses an attention mechanism to focus on the important predicates between $s$ and $o$. Refer to Figure~\ref{fig:modelstructure} for an illustration. The final score of $(s,p,o)$ is given by:
\begin{align}
\label{eq:Hybrid}
score(s,p,o) &= f_1(S^{Att}(s,p,o)) * ReLU(\alpha_1) \nonumber \\
      &+ f_2(S^{ENE}_{subj}(s,p)) * ReLU(\alpha_2)  \nonumber \\
      &+ f_3(S^{ENE}_{obj}(p,o)) * ReLU(\alpha_3)  \nonumber
\end{align}
where $f_i(X) = \sigma(a_iX+b_i)$ normalizes different scores to a comparable distribution. $ReLU(\alpha_i)$ enforces non-negative weights that allow scores to only contribute to the final model without canceling each other. $a_i, b_i, \alpha_i$ are free parameters that are learned during the back propagation gradient descent process.

\subsection{Training Process}
Our task is posed as a ranking problem. Given an entity pair, we want the observed KB relations between them to have higher scores than the unobserved ones. Thus, a pair-wise ranking based loss function is used to train our model:
\begin{align}
L(s,p_{pos},p_{neg},o) = \max(0, \gamma &- score(s,p_{pos},o) \nonumber\\
&+ score(s,p_{neg},o)) \nonumber
\end{align}
where $p_{pos}$ refers to a positive relation, $p_{neg}$ refers to a uniformly sampled negative relation, and $\gamma$ is the margin hyper-parameter. We optimize the loss function using Adam~\cite{kingma2014adam}. 
The training process uses early stop according to the validation set.

\subsection{Explicit Argument Type Constraint}
\label{subsec:explicit_type} 
Subject and object argument types of relations help in filtering out a large number of candidate relations that do not meet the argument type, and therefore serve as useful constraints for relation inference. Similar to~\cite{yu2017open}, we identify the subject and object argument type of each relation by calculating its probability of co-occurrence with subject / object entity types. During inference, we select candidate relations by performing a post-processing filtering step using the subject and object's type information when available.

\section{Experiments}
\label{sec:experiment}

\subsection{Data}

\begin{table*}[!htb]
\small
\begin{center}
\begin{tabular}{l|rr|rr}
\toprule
& ReVerb +  & ReVerb +        & Ceres +   & Ceres +   \\
& Freebase  & Freebase(/film) & Freebase(/film) & IMDB \\
\midrule
\multicolumn{5}{c}{Training set}\\
\midrule

\# entity pairs for model training      & 40,878      & 1,102        & 23,389        &  64,539 \\
\# KB relation types                    & 250         & 64           & 54            & 66\\
\# OpenIE predicate types               & 124,836    & 35,366     & 124           & 178\\
\midrule
\multicolumn{5}{c}{Test set}\\
\midrule
\# test triples                    & 4938 & 402 & 986 & 998 \\
Avg./Med. \# text edges per entity pair   & 1.74 / 1   & 1.49 / 1    & 1.35 / 1     &  1.23 / 1\\
Avg./Med. \# edges for each subj   & 95.71 / 9  & 100.27 / 23  & 48.80 / 44   &  121.06 / 110  \\
Avg./Med. \# kb edges for each subj   & 61.41 / 4  & 8.30 / 6  & 7.00 / 7   &  30.77 / 29  \\
Avg./Med. \# edges for each obj    & 699.89 / 62  & 24.81 / 8   & 558.33 / 9    &  775.74 / 12  \\
Avg./Med. \# kb edges for each obj   & 325.70 / 23  & 10.11 / 6  & 340.96 / 3   &  606.31 / 6  \\

\bottomrule
\end{tabular}
\end{center}
\vspace*{-1em}
\caption{\label{tab:statistics} Data Statistics (Avg: Average, Med: Median).}
\end{table*}

\begin{table*}[htbp]
\small
\begin{center}
\begin{tabular}{l|r|r|r|r} 
\toprule
  Models & ReVerb +  & ReVerb + & Ceres + & Ceres + \\
  & \ \ \ \ \ \ \ \ \  Freebase  & Freebase(/film) & Freebase (/flim)&\ \ \ \ \ \ \ \ \ \ \  IMDB\\
\midrule
$P(p|\overline p')$ (similar to PMI~\cite{angeli2015leveraging}) & 0.412  &0.301 &0.507 & 0.663\\
$P(p|s,\overline p',o)$ & 0.474 & 0.317 & 0.627 & 0.770 \\
\midrule
E-model~\cite{riedel2013relation}  & 0.215 & 0.156 & 0.431 & 0.506 \\
ENE  & 0.479 & 0.359& 0.646 & 0.808  \\
\midrule
Rowless with MaxR~\cite{verga-neelakantan-mccallum:2017:EACLlong} & 0.318 & 0.285 & 0.481 & 0.659 \\
Rowless with Query Attn.~\cite{verga-neelakantan-mccallum:2017:EACLlong}   & 0.326 & 0.278 & 0.512 & 0.695 \\
\midrule
OpenKI with MaxR & 0.500 & \textbf{0.378} & 0.649 & 0.802 \\
OpenKI with Query Att. & 0.497 &  0.372 & \textbf{0.663} & 0.800 \\
OpenKI with Neighbor Att.  & 0.495 & 0.372 & 0.650 & 0.813\\
OpenKI with Dual Att.  & \textbf{0.505} & 0.365 & 0.658 & \textbf{0.814}\\
\bottomrule

\end{tabular}
\end{center}
\vspace*{-1em}
\caption{\label{tab:perf} Mean average precision (MAP) of different models over four data settings.}
\vspace*{-1em}
\end{table*}


We experiment with the following OpenIE datasets and Knowledge Bases.

\noindent (i) Ceres~\cite{lockard-naacl2019} works on semi-structured web pages (e.g., IMDB) and exploits the DOM Tree 
and XPath~\cite{DBLP:conf/edbtw/OlteanuMFB02} structure in the page to extract triples like {\em (Incredibles 2, Cast and Crew, Brad Bird)} and {\em (Incredibles 2, Writers, Brad Bird)}. 
We apply Ceres on the SWDE~\cite{hao2011one} movie corpus to generate triples. We align these triples to two different knowledge bases: (i) IMDB and (ii) subset of Freebase with relations under {\em /film} domain.  The average length of text predicates is $1.8$ tokens for Ceres extractions.

\noindent (ii) ReVerb~\cite{DBLP:conf/emnlp/FaderSE11} works at sentence level and employs various syntactic constraints like  part-of-speech-based regular expressions and lexical constraints to prune incoherent and uninformative extractions. 
We use 3 million ReVerb extractions from ClueWeb where the subject is already linked to Freebase~\cite{lin2012entity}~\footnote{Extractions are downloadable at \url{http://knowitall.cs.washington.edu/linked\_extractions/}}. We align these extractions to (i) entire Freebase and (ii) subset of Freebase with relations under {\em /film} domain. The average length of text predicates is $3.4$ tokens for ReVerb extractions.

In order to show the generalizability of our approach to traditional (non OpenIE) corpora, we also perform experiments in the New York Times (NYT) and Freebase dataset~\cite{riedel2010modeling}, which is a well known benchmark for distant supervision relation extraction. We consider the sentences (average length of $18.8$ tokens) there to be a proxy for text predicates. These results are presented in Section~\ref{sec:nyt}. 



\noindent {\bf Data preparation}: We collect all entity mentions $M$ from OpenIE text extractions, and all candidate entities $E_{KB}$ from KB whose name exists in $M$. We retain the sub-graph $G_{KB}$ of KB triples where the subject and object belongs to $E_{KB}$. 
Similar to~\cite{riedel2013relation}, we use string match to collect candidate entities for each entity mention. For each pair of entity mentions, we link them if two candidate entities in $E_{KB}$ share a relation in KB. Otherwise, we link each mention to the most common candidate. For entity mentions that cannot be linked to KB, we consider them as new entities and link together mentions that share same text .

For validation and test, we randomly hold-out a part of the entity pairs from $G_{KB}$ where text predicates are observed. Our training data consists of the rest of $G_{KB}$ and all the OpenIE text extractions. 
In addition, we exclude direct KB triples from training where corresponding entity pairs appear in the test data (following the data setting of~\cite{toutanova2015representing}).
Table~\ref{tab:statistics} shows the data statistics~\footnote{Our datasets with train, test, validation split are downloadable at \url{https://github.com/zhangdongxu/relation-inference-naacl19} for benchmarking.}.

We adopt a similar training strategy as Universal Schema for the Ceres dataset -- that not only learns direct mapping from text predicates to KB relations, but also clusters OpenIE predicates and KB relations by their co-occurrence. However, for the ReVerb data containing a large number of text predicates compared to Ceres, we only  learn the direct mapping from text predicates to KB relations that empirically works well for this dataset.



\subsection{Verifying Usefulness of Neighborhood Information: Bayesian Methods}
\label{sec:sanity_check}






To verify the usefulness of the entity's neighborhood information, we devise simple Bayesian methods as baselines.
The simplest method counts the co-occurrence of text predicates and KB relations (by applying Bayes rule) to find the conditional probability $P(p|\overline p')$ of a target KB relation $p$ given a set of observed text predicates $\overline p'$. This performs relation inference at {\em predicate-level}.

Then, we can include the entity's relational neighbors in the Bayesian network by adding the neighboring predicates and relations of the subject (given by $p_s^N$) and object (given by $p_o^N$) to find $P(p|s,\overline p',o)$, which performs relation inference at the {\em instance-level}. The graph structures of these three Bayesian methods are shown in Figure~\ref{fig:conditional}.
For detailed formula derivation, please refer
to {\em Appendix}~\ref{app:derivation}.

\begin{figure}[h]
	\centering
	\includegraphics[width=0.25\textwidth]{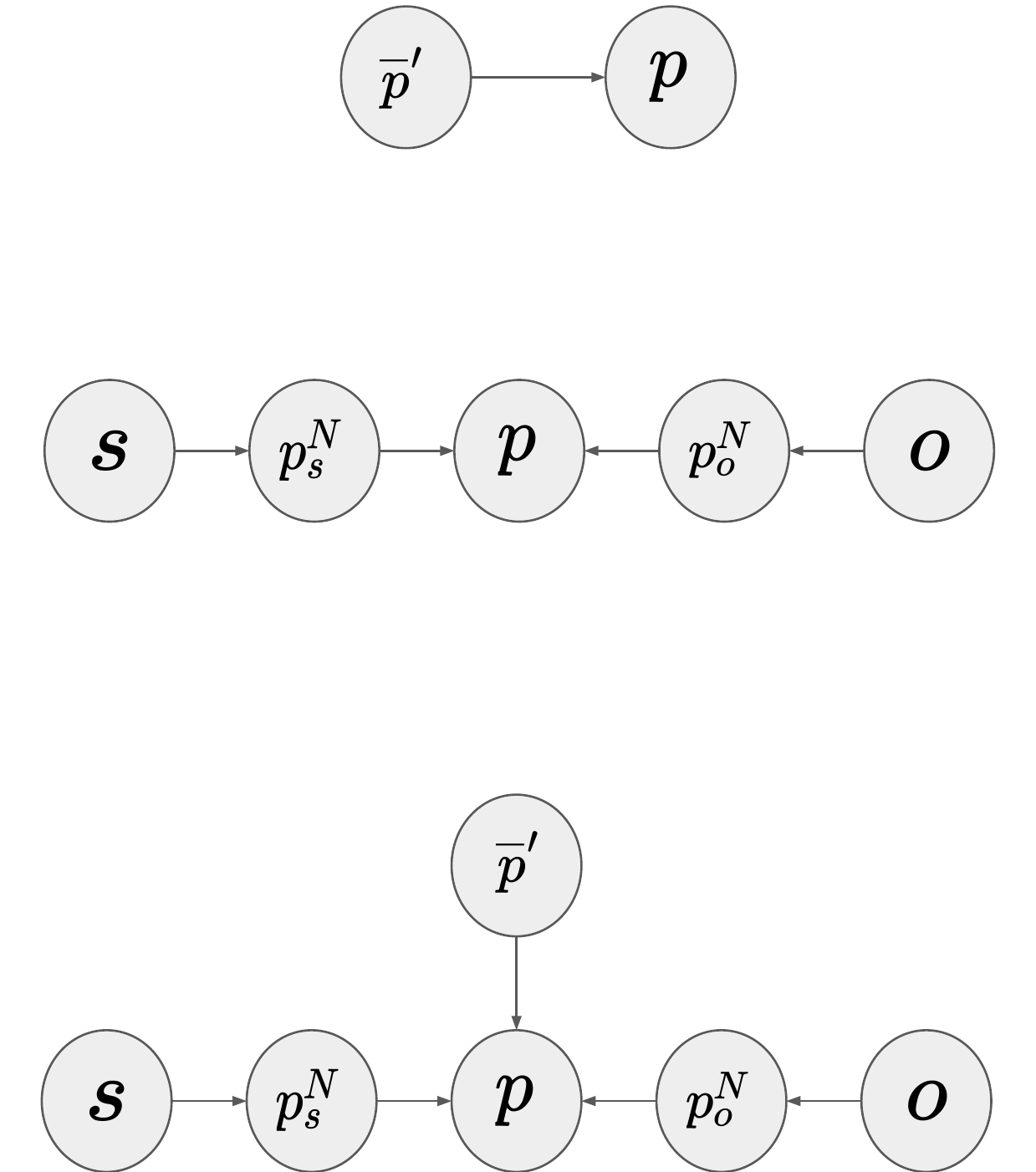}
	\caption{Structures of $P(p|\overline p')$, $P(p|s,o)$, $P(p|s,\overline p',o)$ are listed from top to bottom.}
	\label{fig:conditional}
	\vspace{-1em}
	\centering
\end{figure}

\subsection{Baselines and Experimental Setup} 

\newcite{angeli2015leveraging} employ point-wise mutual information (PMI) between target relations and observed predicates to map OpenIE predicates to KB relations. This is similar to our Bayes conditional probability $P(p|\overline p')$. This baseline operates at {\em predicate-level}. 
To indicate the usefulness of entity neighborhood information, we also compare with $P(p|s,\overline p', o)$ as mentioned in Section~\ref{sec:sanity_check}. For the advanced embedding-based baselines, we compare with the E-model and the Rowless model (with MaxR and query attention) introduced in Section~\ref{sec:background}. 

\noindent {\bf Hyper-parameters:} In our experiments, we use 25 dimensional embedding vectors for the Rowless model, and 12 dimensional embedding vectors for the E- and ENE models. We use a batchsize of 128, and 16 negative samples for each positive sample in a batch. Due to memory constraints, we sample at most 8 predicates between entities and 16 neighbors for each entity during training. We use $\gamma=1.0$ and set the learning rate to 5e-3 for ReVerb and 1e-3 for Ceres datasets.

\noindent{\bf Evaluation measures:} Our task is a multi-label task, where each entity pair can share multiple KB relations. Therefore, we consider each KB relation as a query and compute the Mean Average Precision (MAP) -- 
where entity pairs sharing the query relation should be ranked higher than those without the relation. 
In Section~\ref{subsec:results} we report MAP statistics for the 50 most common KB relations for ReVerb and Freebase dataset, and for the 10 most common relations in other domain specific datasets. The left out relations involve few triples to report any significant statistics. 
We also report the area under the precision recall curve (AUC-PR) for evaluation in Section~\ref{sec:nyt}.

\subsection{Results}
\label{subsec:results}

Table~\ref{tab:perf} shows that the overall results. OpenKI achieves significant performance improvement over all the baselines. Overall, we observe 33.5\% MAP improvement on average across different datasets.

From the first two rows of Table~\ref{tab:perf}, we observe the performance to improve as we incorporate neighborhood information into the Bayesian method. This depicts the strong influence of the entity's neighboring relations and predicates for relation inference.

The results show that our Entity Neighbor Encoder (ENE) outperforms the E-Model significantly. This is because the majority of the entity pairs in our test data have {\em at least one unseen entity} (refer to Table~\ref{tab:unseen_stats}), which is very common in the OpenIE setting. The E-model cannot handle unseen entities because of its modeling of entity-specific parameters.  This demonstrates the benefit of encoding entities with their neighborhood information (KB relations and text predicates) rather than learning entity-specific parameters. 
Besides, ENE outperforms the Rowless Universal Schema model, which does not consider any information surrounding the entities. This becomes a disadvantage in sparse data setting where only a few predicates are observed between an entity pair. 

Finally, the results also show consistent improvement of OpenKI model over only-Rowless and only-ENE models. This indicates that the models are complementary to each other. We further observe significant improvements by applying different attention mechanisms over the OpenKI MaxR model -- thus establishing the effectiveness of our attention mechanism.

\noindent \textbf{Unseen entity:} Table~\ref{tab:unseen_stats} shows the data statistics of unseen entity pairs in our test data. The most common scenario is that only one of the entity in a pair is observed during training, where our model benefits from the extra neighborhood information of the observed entity in contrast to the Rowless model. 

\begin{table}[h]
\small
\begin{center}
\begin{tabular}{l|ccc} 
\toprule
Dataset &  Both & One & Both \\
 &seen & unseen & unseen \\
\midrule
ReVerb + Freebase & 864 & 3232  & 842  \\
ReVerb + Freebase(/film) & 27 & 147 &  228   \\
Ceres + Freebase(/film) & 383  & 561  & 42  \\
Ceres + IMDB  & 462 & 533 & 3  \\
\bottomrule
\end{tabular}
\end{center}
\vspace{-1em}
\caption{\label{tab:unseen_stats} Statistics for unseen entities in test data. ``Both seen'' indicates both entities exist in training data; ``One unseen'' indicates only one of the entities in the pair exist in training data; ``Both unseen'' indicates both entities were unobserved during training.}
\vspace{-1em}
\end{table}

\begin{table}[h]
\small
\begin{center}
\begin{tabular}{l|c|c} 
\toprule
Models  &  All data & At least one seen \\
\midrule
Rowless Model  &  0.278 & 0.282  \\ 
OpenKI with Dual Att. & 0.365   &  0.419  \\
\bottomrule

\end{tabular}
\end{center}
\vspace{-1em}
\caption{\label{tab:unseen} Mean average precision (MAP) of Rowless and OpenKI on ReVerb + Freebase (/film) dataset.}
\vspace{-1em}
\end{table}

Table~\ref{tab:unseen} shows the performance comparison on test data where at least one of the entity is known at test time. 
We choose ReVerb+Freebase(/film) for analysis because it contains the largest proportion of test triples where both entities are unknown during training. 
From the results, we observe that OpenKI outperforms the Rowless model by 48.6\% when at least one of the entity in the triple is observed during training. Overall, we obtain 31.3\% MAP improvement considering all of the test data. This validates the efficacy of encoding entity neighborhood information where at least one of the entities is known at test time. In the scenario where both entities are unknown at test time, the model falls back to the Rowless setting.




\begin{table}[h]
\small
\begin{center}
\begin{tabular}{l|l} 
\toprule
Models    & MAP          \\

\midrule
Rowless Model          &  0.695    \\
+Type Constraint     &  0.769  \ \ $ \uparrow$ 10.6\%  \\
\midrule
ENE Model                    &  0.808  \\
+Type Constraint     &  0.818  \ \  $\uparrow$ 1.2\% \\
\midrule
OpenKI with Dual Att.        &   0.814  \\
+Type Constraint      &  \textbf{0.828}   \ \  $\uparrow$ 1.7\%\\
\bottomrule

\end{tabular}
\end{center}
\vspace{-1em}
\caption{\label{tab:type_constraint} MAP improvement with argument type constraints on Ceres + IMDB dataset.}
\vspace{-1em}
\end{table}


\noindent \textbf{Explicit Argument Type Constraint:}
As discussed in Section~\ref{subsec:explicit_type}, incorporating explicit type constraints can improve the model performance. However, entity type information and argument type constraints are not always available especially for new entities. Table~\ref{tab:type_constraint} shows the performance improvement of different models with entity type constraints. We observe the performance improvement of the ENE model to be much less than that of the Rowless model with explicit type constraint.  This shows that the ENE model already captures soft entity type information while modeling the neighborhood information of an entity in contrast to the other methods that require explicit type constraint.



\subsection{Results on NYT + Freebase Dataset}
\label{sec:nyt}

Prior works~\cite{surdeanu2012multi,zeng2015distant,lin2016neural,qin2018robust} on distantly supervised relation extraction performed evaluations on the New York Times (NYT) + Freebase benchmark data  developed by~\newcite{riedel2010modeling}\footnote{This data can be downloaded from \url{http://iesl.cs.umass.edu/riedel/ecml/}}. The dataset contains sentences whose entity mentions are annotated with Freebase entities as well as relations. The training data consists of sentences from articles in 2005-2006 whereas the test data consists of sentences from articles in 2007. There are 1950 relational facts in our test data\footnote{Facts of `NA' (no relation) in the test data are not included in the evaluation process.}. In contrast to our prior experiments in the semi-structured setting with text predicates, in this experiment we consider the sentences to be a proxy for the text predicates.

\begin{table}[h]
\small
\begin{center}
\begin{tabular}{l|r} 
\toprule
Models    & AUC-PR \\
\midrule
PCNN + MaxR~\cite{zeng2015distant}  & 0.325 \\
PCNN + Att.~\cite{lin2016neural}  & 0.341 \\
\midrule
ENE & 0.421 \\
OpenKI with Dual Att. & \textbf{0.461}\\
\bottomrule

\end{tabular}
\end{center}
\vspace{-1em}
\caption{\label{tab:nyt} Performances on NYT + Freebase data.}
\vspace{-1em}
\end{table}

Table~\ref{tab:nyt} compares the performance of our model with two state-of-the-art works~\cite{zeng2015distant,lin2016neural} on this dataset using AUC-PR as the evaluation metric. 

Overall, OpenKI obtains 35\% MAP improvement over the best performing PCNN baseline. In contrast to baseline models, our approach leverages the neighborhood information of each entity from the text predicates in the 2007 corpus and predicates / relations from the 2005-2006 corpus. This background knowledge contributes to the significant performance improvement.

Note that, our model uses only the graph information from the entity neighborhood and does not use any text encoder such as Piecewise Convolutional Neural Nets (PCNN)~\cite{zeng2015distant}, where convolutional neural networks were applied with piecewise max pooling to encode textual sentences.  This further demonstrates the importance of entity neighborhood information for relation inference. It is possible to further improve the performance of our model by incorporating text encoders as an additional signal. Some prior works~\cite{verga-EtAl:2016:N16-1,toutanova2015representing} also leverage  text encoders for relation inference.


\section{Related Work}
\label{sec:related_work}
\noindent {\bf Relation Extraction:} 
\newcite{mintz2009distant} utilize the entity pair overlap between knowledge bases and text corpus to generate signals for automatic supervision. 
To avoid false positives during training, many works follow the at-least-one assumption, where at least one of the text patterns between the entity pair indicate an aligned predicate in the KB ~\cite{hoffmann2011knowledge,surdeanu2012multi,zeng2015distant,lin2016neural}. These works do not leverage graph information. 
In addition, Universal Schema~\cite{riedel2013relation,verga-neelakantan-mccallum:2017:EACLlong} tackled this task by low-rank matrix factorization.
\newcite{toutanova2015representing} exploit graph information for knowledge base completion. However, their work cannot deal with unseen entities since entities' parameters are explicitly learned during training. 

\noindent {\bf Schema Mapping:}
Traditional schema mapping methods~\cite{rahm2001survey} involve three kinds of features, namely, language (name or description), type constraint, and instance level co-occurrence information. 
These methods usually involve hand-crafted features. In contrast, our model learns 
all the features automatically from OpenIE and KB with no feature engineering. This makes it easy to scale to different domains with little model tuning.
Also, the entity types used in traditional schema mapping is always pre-defined and coarse grained, so cannot provide precise constraint of relations for each entity. Instead, our ENE model automatically learns soft and fine-grained constraints on which relations entities are likely to participate in. It is also compatible with pre-defined type systems.

\noindent {\bf Relation Grounding from OpenIE to KB:} 
Instead of modeling existing schema, open information extraction (OpenIE)~\cite{banko2007open,DBLP:conf/naacl/YatesBBCES07,DBLP:conf/emnlp/FaderSE11,DBLP:conf/emnlp/MausamSSBE12}  regards surface text mentions between entity pairs as separate relations, and do not require entity resolution or linking to KB. Since they do not model KB, it is difficult to infer KB relations only based on textual observations.
\newcite{soderland2013open} designed manual rules to map relational triples to slot types.
~\newcite{angeli2015leveraging} used PMI between OpenIE predicates and KB relations using distant-supervision from shared entity pairs for relation grounding. \newcite{yu2017open} used word embedding to assign KB relation labels to OpenIE text predicates without entity alignment. These works do not exploit any graph information. 


\noindent {\bf Entity Modeling for Relation Grounding:}
People leveraged several entity information to help relation extraction. 
\newcite{guodong2005exploring} employed type information and observed 8\% improvement of F-1 scores.
\newcite{ji2017distant} encoded entity description to calculate attention weights among different text predicates within an entity pair.
However, entity type and description information is not commonly available. Instead, the neighborhood information is easier to obtain and can also be regarded as entities' background knowledge. 
Universal Schema~\cite{riedel2013relation} proposed an E-Model to capture entity type information. However, it can easily overfit in the OpenIE setting with large number of entities and a sparse knowledge graph.

\section{Conclusion}

In this work we jointly leverage relation mentions from OpenIE extractions and knowledge bases (KB) for relation inference and aligning OpenIE extractions to KB. Our model leverages the rich information (KB relations and OpenIE predicates) from the neighborhood of entities to improve the performance of relation inference. This also allows us to deal with new entities without using any entity-specific parameters. We further explore several attention mechanisms to better capture entity pair information. Our experiments over several datasets show 33.5\% MAP improvement on average over state-of-the-art baselines. 

Some future extensions include exploring more advanced graph embedding techniques without modeling entity-specific parameters and using text encoders as additional signals.



\bibliography{naaclhlt2019}
\bibliographystyle{acl_natbib}
\newpage

\newpage
\appendix
\label{sec:appendix}

\section{Derivation of Bayesian Inference Baselines}
\label{app:derivation}
\small
\begin{align}
& P(p|\overline p') = \max_{p' \in R(s,o)} P(p|p') = \max_{p' \in R(s,o)} \frac{P(p,p')}{P(p')} \nonumber \\
& = \max_{p' \in R(s,o)} \frac{\#(p,p')/\#entity\ pair}{\#p'/\#entity\ pair} \nonumber \\
& \approx \max_{p' \in R(s,o)} \frac{\#(p, p')+\Delta}{\#(p')+|R^{text} \cup R^{KB}|\Delta}  \nonumber
\end{align}
\normalsize

$R(s,o)$ is the set of observed predicates between subject $s$ and object $o$. 
$\Delta$ is a smoothing factor which we choose 1e-6 in our implementation. Using conditional independence assumptions (refer to Figure~\ref{fig:conditional}):

\small

\begin{align}
& P(p|s,o) = \frac{P(p,s,o)}{P(s,o)} = \frac{P(p|s)P(s)P(p|o)P(o)}{P(s)P(o)} \nonumber \\
& = \sum_{p_s^N \in R(s,\cdot)} P(p_s^N|s)P(p|p_s^N) \nonumber \\
& \cdot \sum_{p_o^N \in R(\cdot, o)} P(p_o^N|o)P(p|p_o^N) \nonumber
\end{align}

\begin{align}
& P(p|s,
\overline p',o) = \frac{P(p,s,\overline p',o)}{P(s,\overline p',o)} \nonumber \\
&= \frac{P(p|s)P(s)P(p|\overline p')P(\overline p')P(p|o)P(o)}{P(s)P(\overline p')P(o)} \nonumber \\
& = P(p|\overline p')\sum_{p_s^N \in R(s,\cdot)} P(p_s^N|s)P(p|p_s^N) \nonumber \\
& \cdot \sum_{p_o^N \in R(\cdot, o)} P(p_o^N|o)P(p|p_o^N) \nonumber
\end{align}

\normalsize

\end{document}